\documentclass[12pt,a4paper]{article}

\usepackage{graphicx}
\usepackage{dcolumn}
\usepackage{bm}
\usepackage{amssymb}
\usepackage{setspace}
\usepackage{sectsty}

\allsectionsfont{\large \textbf}

\begin{document}
\begin{center}
{\large {\bf Stabilisation of BGK modes by relativistic effects}}\\
\underline{N. J. Sircombe}$^{1,2}$, M. E. Dieckmann$^{2,3}$,\\  P. K. Shukla$^{2}$ , T. D. Arber$^{1}$\\
{{\em$^1$Department of Physics, University of Warwick,\\ Coventry CV4 7AL, UK\\
$^2$Ruhr-University Bochum, Institute of Theoretical Physics IV,\\ NB 7/56, D-44780 Bochum, Germany\\
$^3$Linkoping University, Department of Science and Technology,\\ SE-60174, Norrkoping, Sweden 
}}
\end{center}

   \abstract
   {We investigate the acceleration of electrons via their 
interaction with electrostatic waves, driven by the relativistic 
Buneman instability, in a system dominated by counter-propagating 
proton beams. We observe the growth of these waves and their subsequent 
saturation via electron trapping for a range of proton beam velocities, 
from $0.15c$ to $0.9c$. We can report a reduced stability of the 
electrostatic wave (ESW) with increasing non-relativistic beam velocities 
and an improved wave stability for increasing relativistic beam velocities, 
both in accordance with previous findings. At the highest beam speeds, we 
find the system to be stable again for a period of $\approx 160$ plasma 
periods. Furthermore we observe a, to our knowledge, previously unreported 
secondary electron acceleration mechanism at low beam speeds. We believe 
that it is the result of parametric couplings to produce high phase 
velocity ESW's which then trap electrons, accelerating them to higher 
energies. This allows electrons in our simulation study to achieve the 
injection energy required for Fermi acceleration, for beam speeds as low 
as $0.15c$ in unmagnetised plasma.}
\newpage
\section{Introduction}
{Supernova remnant (SNR) shocks, created by the interaction between an expanding supernova blast shell and the ambient medium, are believed to be a significant source of cosmic rays (Reynolds \cite{Reynolds01}; Lazendic et al. \cite{Lazendic04}), with energies up to $10^{14}$ eV, the knee of the cosmic ray spectrum  (Nagano \& Watson \cite{Nagano00}; Aharonian et al. \cite{Aharonian04}; V\"{o}lk et al. \cite{Volk88}).}

{First order Fermi acceleration has been proposed as a mechanism for the acceleration of electrons to ultra-relativistic velocities in SNR shocks (Bell \cite{Bell78i}; Bell \cite{Bell78ii}; Blandford \& Ostriker \cite{Blandford78}).} Electrons gain energy by repeated crossings of the shock front, and by their scattering off MHD waves on either side of the shock. 
{The orientation of the ambient magnetic field $\vec{B}$ relative to the shock normal has implications for the efficiency of electron acceleration (Galeev \cite{Galeev84}). We consider the case where the ambient field is orthogonal to the shock normal, this geometry provides an efficient electron acceleration mechanism for high Mach number shocks (Treumann \& Terasawa \cite{Treumann}). First order} Fermi acceleration at such shocks requires a seed population of electrons that have Larmor radii comparable to the shock thickness. Since the shock thickness is, for perpendicular shocks, 
of the order of the ion Larmor radius, the electrons of this seed 
population must have mildly relativistic initial speeds. 
A kinetic energy comparable to 100 keV is believed to be sufficient 
(Treumann \& Terasawa \cite{Treumann}). Such electrons are 
unlikely to be present, neither in the interstellar medium (ISM) nor 
in the stellar wind of the progenitor star of the supernova, {but may be created by a pre-acceleration mechanism at the SNR shock front. This pre-acceleration, commonly referred to as the injection problem, is not well understood.}

As the shock front of a supernova remnant (SNR) expands into the ISM, 
it reflects a substantial fraction of the ISM ions as observed in 
simulations (Shimada \& Hoshino \cite{Shimada3}; Schmitz et al. 
\cite{Schmitz}) {and in-situ at the Earth's bow shock (Eastwood 
et al. \cite{Foreshock}).} If the shock normal is quasi-perpendicular 
to the magnetic field in a high Mach number shock, as many 
as 20\% of the ions can be reflected {(Sckopke et al. \cite{Scopce83}; 
Galeev \cite{Galeev84}; Lembege \& Savoini \cite{Lembege92}; Lembege et al. 
\cite{Lembege04})}. The reflected ions form a beam that can reach a peak 
speed comparable to twice the shock speed in the ISM frame of reference 
{as, for example, discussed by McClements et al. (\cite{KenPaper}). 
The shock-reflected plasma particles are a source of free energy,
similar to the shock-generated cosmic rays (Zank et al. \cite{Zank90}) which are
also thought to heat the inflowing plasma. However, the shock
reflected ion beam is considerably more dense than the cosmic
rays and each particle carries less energy. The developing
plasma thermalisation mechanisms are thus likely to be different.} 
Since binary collisions between charged particles in the dilute ISM plasma 
are negligible, these ion beams relax by their interaction with electrostatic 
waves 
and electromagnetic waves. 
{In what follows, we focus on the interaction of electrons with 
high-frequency electrostatic waves (ESWs) that are driven by two-stream 
instabilities.}

Recent {particle-in-cell (PIC)} simulation studies (Shimada \& Hoshino 
\cite{Shimada2}; Shimada \& Hoshino \cite{Shimada1}; Dieckmann et al. 
\cite{MarkPoP1}; Dieckmann et al. \cite{MarkPoP2}; McClements et al. 
\cite{KenPRL}) have examined these mechanisms with a particular focus on 
how and up to what energies the ESWs driven by non-relativistic or mildly 
relativistic ion beams can accelerate the electrons in the foreshock region. 

  \begin{figure*}
   \centering
   \includegraphics[width=12cm]{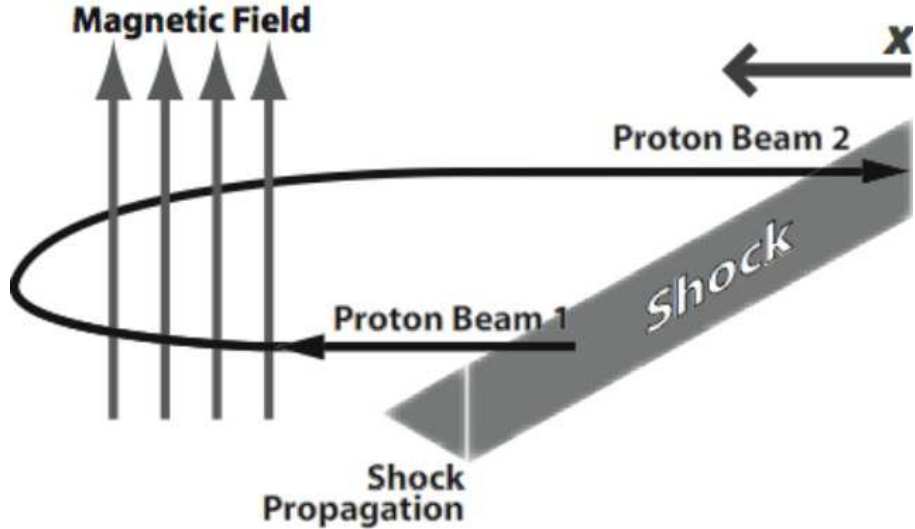}
   	\caption{As the SNR shock expands it reflects a fraction of the ISM 
protons. These protons move back into the upstream region and form beam 1. 
The upstream $\vec{B}$ rotates beam 1 which returns as beam 2. The simulation 
box covers a one dimensional region of {{\em x}} in front of the shock sufficiently small that we can 
assume it to be spatially homogeneous and one dimensional.}
	\label{Fig1}
    \end{figure*}
The maximum energy the electrons can reach by such wave-particle interactions, depends on the life-time of the 
saturated ESW and on the strength and the orientation of $\vec{B}$. Initially 
a stable non-linear wave known as BGK mode (Bernstein et al. \cite{Bernstein}; 
Manfredi \cite{Manfredi}; Brunetti et al. \cite{Brunetti}) develops if the 
plasma is unmagnetised or if the magnetic field is weak (McClements et al. 
\cite{KenPRL}; Dieckmann et al. \cite{MarkIEEE}; Eliasson, Dieckmann \& Shukla 
\cite{Eliasson05}). Such modes are associated with phase space holes - islands of trapped electrons. These BGK modes are destabilised by the 
sideband instability, a resonance between the electrons that 
oscillate in the ESW potential and secondary ESWs (Kruer et al. \cite{Kruer}; 
Tsunoda \& Malmberg \cite{Tsunoda}; Krasovsky \cite{Krasovsky}). This 
resonance transfers energy from the trapped electrons to the secondary 
ESWs. The initial BGK mode collapses, once these secondary ESWs grow to 
an amplitude that is comparable to that of the initial wave.

{Many previous simulations of two-stream instabilities in the 
context of electron injection and of shocks have employed 
PIC simulation codes, which suffer from high noise levels 
(Dieckmann et al. \cite{Noise}) and from a dynamical range for the 
plasma phase space distribution that is limited by the number of 
computational particles. It is thus possible that certain instabilities 
can not develop, due to a lack of computational particles in the relevant 
phase space interval, or that the phase space structures (e.g. BGK modes) 
are destabilised by the noise (Schamel \& Korn \cite{Schamel}) and may benefit from a plasma model based on the direct solution of the Vlasov equation. 
Such effects have been demonstrated 
in the non-relativistic limit, for example by Eliasson, Dieckmann \& Shukla 
(\cite{Eliasson05}) and Dieckmann et al. (\cite{MarkPRL}), in which the 
results computed by PIC codes have been compared to those of Vlasov codes.  
The particle species in a Vlasov code are represented by a continuous distribution function, conventionally evolved on a fixed Eulerian grid, rather than 
by simulation macro-particles. The comparison of results from these two methodologies has shown significant differences 
in the life-time of the BGK modes both for unmagnetized and magnetized plasma.} 

SNR shocks typically expand into the ISM at speeds ranging between a 
few and twenty percent of $c$ (Kulkarni et al. \cite{Kulkarni1998}). The 
shock-reflected ions can thus reach a speed of $v_b \approx 0.4c$ if the 
reflection is specular (McClements et al. \cite{KenPaper}) or even higher 
speeds if we take into account shock surfing acceleration (Ucer \& Shapiro 
\cite{Shapiro1}; Shapiro \& Ucer \cite{Shapiro2}). The proton-beam driven 
ESWs have a phase speed similar to $v_b$ (Buneman \cite{Buneman}; 
Thode \& Sudan \cite{Thode}). Since the ISM electrons can be accelerated 
to a maximum speed well in excess of the phase speed of the ESW 
(Rosenzweig \cite{Rosenzweig}), we must consider relativistic modifications 
of the life-time of the BGK modes in SNR foreshock plasma. It has been found 
by Dieckmann et al. (\cite{MarkPoP2}) that the BGK mode is stabilised if the 
phase speed of the ESW is relativistic in the ISM frame of reference. The 
likely reason is that the change in the relativistic electron mass introduces 
a strong dependence of the electron bouncing frequency on the electron speed 
in the ESW frame of reference. This decreases the coherency with which the 
electrons interact with the secondary ESWs and thus the efficiency of the 
sideband instability. The stabilization is clearly visible in 
PIC simulations for ESWs moving with a phase speed of 
$0.9c$ (Dieckmann et al. \cite{MarkPoP2}). For speeds $v_b < 0.7c$ the 
relativistic modifications of the BGK mode stability have been small in 
the PIC simulations (Dieckmann et al. \cite{MarkPoP2}). The better 
representation of the phase space density afforded by a Vlasov simulation 
may, however, yield a different result {and it requires a further 
examination}. 



{We thus focus in this work on the modelling of electrostatic instabilities by means of relativistic Vlasov simulations and we assess the impact of these instabilities as a potential pre-acceleration mechanism. To this end, we neglect magnetic field effects and consider only the one-dimensional electrostatic system, which is equivalent to the approach taken by Dieckmann et al. (\cite{MarkPoP2}) but at a much larger dynamical range for the plasma phase space distribution. This description serves as simple model for the stability of ESWs by which we identify similarities and differences between the results provided by PIC and Vlasov simulations. Future work will expand the simulations to include magnetic fields, which introduces electron surfing acceleration (ESA) (Katsouleas \& Dawson \cite{Katsouleas}; McClements et al. \cite{KenPRL}) and stochastic particle orbits (Mohanty \& Naik \cite{ObliqueMode}), and multiple dimensions.}

{More specifically,} we examine by relativistic electrostatic 
Vlasov simulations (Arber \& Vann \cite{Arber02}; Sircombe et al. 
\cite{Sircombe05}) how the BGK mode life-time and its collapse in an 
unmagnetised plasma depend on $v_b$ and thus on the phase speed of the 
ESW. The purpose is twofold. First, we extend the comparison of results 
provided by PIC and by Vlasov codes beyond the nonrelativistic 
regime in Dieckmann et al. (\cite{MarkPRL}). We perform Vlasov 
simulations for initial conditions that are identical to those in 
Dieckmann et al. (\cite{MarkPoP2}) where a PIC simulation code 
(Eastwood \cite{Eastwood91}) has been used. We find a good agreement 
of the results of the relativistic Vlasov code and the PIC code for 
relativistic $v_b$ and with the corresponding result provided by
the nonrelativistic Vlasov code (Eliasson \cite{Eliasson02};
Dieckmann et al. \cite{MarkPRL}). We thus bring forward further evidence 
for both, an increasing destabilisation of BGK modes for increasing 
nonrelativistic phase speeds of the wave and a stabilisation for 
increasing relativistic phase speeds.
Secondly we want to exploit the much higher dynamical range of
Vlasov simulations to examine the wave spectrum and the particle 
energy spectrum that we obtain by the considered nonlinear 
interactions. We find that the collapse of the BGK modes
couples energy to three families of waves. Firstly a continuum of 
electrostatic waves that move with approximately the beam speed. 
These waves are connected to the turbulent electron phase space
flow. Secondly we find for high beam speeds the growth of
quasi-monochromatic modes with frequency comparable
to the Doppler-shifted bouncing frequency of the trapped electrons
in the wave potential. These waves would be sideband modes (Kruer et 
al. \cite{Kruer}). Thirdly we find waves that do not have a clear 
connection to any characteristic particle speed. We believe that 
these modes are produced by parametric instabilities. These modes 
can reach phase speeds well above the maximum speed the initial 
trapped electron population reaches, and they grow to amplitudes at 
which they can trap electrons, i.e. a BGK mode cascade to high speeds 
develops. By this trapping cascade, the electrons can reach momenta 
well in excess of those reported previously (Dieckmann et al. 
\cite{MarkPoP1}; Dieckmann et al. \cite{MarkPoP2}). This result
would imply that the ion beams, that are reflected by shocks that
expand at speeds comparable to SNR shocks, can accelerate electrons
to energies in excess of 100 keV, at which they can undergo Fermi 
acceleration to higher energies. 

\section{The physical model, the linear instability and the 
simulation setup}

\subsection{Physical model}
We consider a small interval of the ISM plasma just ahead of the SNR 
shock (see Fig. \ref{Fig1}) and we treat the ISM plasma as an electron 
proton plasma with a spatially homogeneous Maxwellian velocity 
distribution. We place our simulation box close to the SNR shock, so that 
the shock-reflected ISM protons can cross it. The protons, that have just 
been reflected, constitute beam 1, beam 2 represents the protons that 
return to the shock after they have been rotated by the global foreshock 
magnetic field. Our model is in line with that discussed, for example, 
by McClements et al. (\cite{KenPaper}) and solved numerically, e.g. for 
unmagnetized plasma by Dieckmann et al. (\cite{MarkPoP1,MarkPoP2,MarkPRL}) 
and for magnetized plasma by McClements et al. (\cite{KenPRL}) and by 
Shimada \& Hoshino (\cite{Shimada1}). 


We 
set $\vec{B} = 0$, {and thus, exclude electromagnetic
instabilities, e.g. Whistler waves (Kuramitsu \& Krasnoselskikh 
\cite{Volodya}), MHD waves, and electrostatic waves in magnetised 
plasma, e.g. electron cyclotron waves. However, by this choice we 
decouple the development of competing instabilities and we can 
consider them separately. In this work we focus on ESWs and the 
nonlinear BGK modes, which are important phase space structures in 
the foreshocks of Solar system plasma shocks (Treumann \& Terasawa \cite{Treumann}).}

\begin{figure}
   \centering
   \includegraphics[width=12cm]{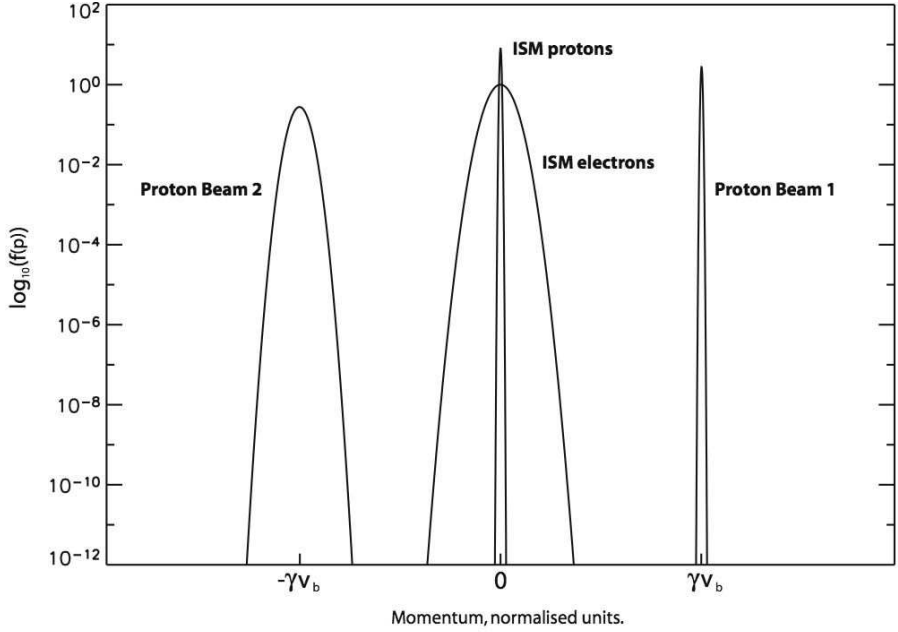}
   \caption{The initial velocity distribution showing the ISM electrons 
and protons, proton beam 1 and proton beam 2. The expanding SNR shock 
reflects a fraction of the ISM protons which form beam 1. The upstream 
$\vec{B}$ rotates beam 1 to create beam 2. Since the beam has been 
scattered by ESWs on its path through the foreshock, its thermal spread 
has increased. Both beams can grow ESWs with a speed similar to $v_b$ 
which saturate by their interaction with the ISM electrons. By assuming 
that the plasma parameters do not change over the small box size we 
consider, we can represent this plasma by the 4 spatially homogeneous 
species shown.}\label{Fig2}
    \end{figure}

The system is{, with the choice $\vec{B}=0$,} 
suitable for modelling with an electrostatic and relativistic
Vlasov-Poisson solver.
The size of the simulation box is small compared to the distance 
across which the beam parameters change.
Thus we can take periodic boundary conditions for the 
simulation and spatially homogeneous Maxwellian distributions for 
both beams. Both beams have the same mean speed modulus, $|v_b|$, 
but move into opposite directions which gives a zero net current 
in the simulation box. We take a higher temperature for beam 2 
than for beam 1 to reflect the scattering of the beam protons as 
they move through the foreshock. We show the velocity distributions 
in Fig. \ref{Fig2}

{The choice of $\vec{B}=0$ is a critical limitation of the model. The initial conditions described above could not fully describe a perpendicular shock and with $\vec{B} = 0$ one could not account for the presence of a returning proton beam. We assume that, while the field is neglected over the simulation box, the field outside the box is sufficient to produce the proton beam structure described.
Therefore, the results presented here are not directly applicable to the foreshock dynamics of high Mach number shocks. However, they provide an overview of processes that require a large dynamical
range for the plasma phase space distribution and which may, therefore, 
not have been observed in previous PIC simulations. Our system is applicable to
parts of the foreshock of perpendicular shocks, where
magnetic field fluctuations (Jun \& Jones \cite{Jun99}) cause the
magnetic field to vanish or to be beam aligned. It may also apply to 
the field aligned ion beams that are observed in the foreshock region 
of the Earth's bow shock (Eastwood et al. \cite{Foreshock}) and which
may be present also at SNR shocks. Here, the second proton beam in our 
initial conditions takes the role of the return current in the plasma 
which is typically provided by all plasma species (Lovelace \& Sudan 
\cite{Lovelace}). Interactions between two BGK modes 
in unmagnetized plasma affect primarily the velocity interval that is 
confined by the phase speeds of the two waves (Escande \cite{Escande}). 
Our results, which focus on the developing high energy tails of the plasma 
distribution, may thus not depend on the exact setting of the initial 
return current and may be more universally applicable.}

\subsection{Linear instability}
We introduce the plasma frequency of each species $i$ as $\omega_{p,i} = 
{(e^2 n_i / m_i \epsilon_0)}^{1/2}$ where $e$, $n_i$, $m_i$ and
$\epsilon_0$ are the magnitude of the elementary charge, the number 
density of species $i$ in the rest frame of the species, the particle 
mass of species $i$ and the dielectric constant. In what follows we 
define all $\omega_{p,i}$ in the box frame of reference. Species 1 
and 2 are the ISM electrons and protons. The species 3 is beam 1 and 
species 4 is beam 2. The large inertia of species 2 compared to 
species 1 implies, that its contribution to the linear dispersion 
relation can be neglected. The linear dispersion relation then becomes 
in the cold plasma limit
\begin{equation}
\frac{\omega_{p,1}^2}{\omega^2} + \frac{\omega_{p,3}^2}{\gamma^3 (v_b)
{(\omega - v_b k )}^2}+ \frac{\omega_{p,4}^2}{\gamma^3 (v_b)
{(\omega + v_b k )}^2}-1 = 0
\label{Eq1}
\end{equation}
During the linear growth phase, each proton beam will grow an ESW by
its streaming relative to the electrons. Both waves are well separated 
in their phase speed and the linear dispersion relation can be solved 
separately for each ESW by neglecting either the term with $\omega_{p,3}$
or $\omega_{p,4}$ in Eq. \ref{Eq1} (Buneman \cite{Buneman}; Thode \& 
Sudan \cite{Thode}). Its frequency in the box frame 
is $\omega_u \approx \omega_{p,1}$, its wave number is $k_u \approx 
\omega_{p,1} v_b^{-1}$ and its growth rate is $\Omega \approx 
{(3\sqrt{3} \omega_{p,3}^2 \omega_{p,1}/16)}^{1/3} /\gamma (v_b)$.
We take a number density ratio of $n_3 / n_1 = 0.2 \gamma (v_b)$
 and set $n_4=n_3$ and $n_2 = n_1 - n_3 - n_4$ in the box frame of 
reference, which is representative for a shock with a high Mach 
number (Galeev \cite{Galeev84}). This density ratio gives $\omega_{p,1} 
/ \omega_{p,3} \approx 96 {\gamma (v_b)}^{-0.5}$ and a growth rate of 
$\Omega \approx 0.033 / {\gamma (v_b)}^{2/3}$. The growth rate is 
reduced by increasing temperatures of the plasma.
The wave length of the most unstable ESW is $\lambda_u = 2\pi / k_u$. 
The sideband instability couples energy to modes with $k \le k_u$, 
for non-relativistic phase speeds of the wave (Krasovsky \cite{Krasovsky}). 
We thus set the simulation box length 
to $L = 2\lambda_u$ to resolve more than one unstable sideband mode,
which leads to a wave collapse (Dieckmann et al. \cite{MarkPoP1}). 

This $L$ is short compared to the typical size of the foreshock which 
justifies our spatially homogeneous initial conditions and periodic 
boundary conditions. We use the amplitudes of the initial ESW and
that of the sideband modes as indicators for the wave collapse.
We define $l$ as the index of the simulation cell, 
$x_l = l\Delta x$, $k_j = 2\pi j / N_x \Delta x$ and $N_x$, 
as the number of simulation cells in $x$. The index $m$ refers
to the data time step $m\Delta t$, where $\Delta t$ is the time interval 
between outputs rather than the simulation time step. We Fourier Transform 
the spatio-temporal ESW field $E(x,t)$ as 
\begin{equation}
E(k_j,t_m) = \left [ {N_x}^{-1} |\sum_{l=1}^{N_x} E(x_l,t_m) \exp{(ik_jx_l)}| 
\right ] \label{Eq2}
\end{equation}
The amplitude of the ESW with $k=k_u$ at the simulation time $t_m$ is then 
given by $E(k_2,t_m)$.

To analyse the nonlinear and time dependent processes developing after
the saturation of the ESW, we introduce a Window Fourier Transform. We define 
the window size, in time steps, as $N_t$ and we introduce $\omega_n = n \Delta 
\omega = n (2\pi / N_t \Delta t)$ as the frequency defined in a Fourier time 
window with a size $N_t \Delta t$. This frequency is limited by the sampling 
theorem to $-\pi / \Delta t < \omega_n \le \pi / \Delta t$. The Window 
Fourier transform can be written as
\begin{eqnarray}
W_1(k_j,t_m) & = & {N_x}^{-1} \sum_{l=1}^{N_x} E(x_l,t_m) \exp{(ik_jx_l)} \nonumber \\
W_2(\omega_n,t_s,k_c) & = & {N_t}^{-2} 
{\left | {\sum_{l=s-N_t / 2 + 1}^{s+N_t/2} W_1(k_c,t_l)} \exp{(i\omega_nt_l)} 
\right |}^2 \label{Eq3}
\end{eqnarray}
where the Fourier transformed time interval is small compared to the 
total length of the time series, which is $19 N_t\Delta t$, and where 
$k_c$ is a fixed wave number.

\subsection{Numerical Simulation}

In the absence of a magnetic field the one dimensional relativistic 
Vlasov-Poisson system of electrons and protons is given by the Vlasov 
equation for the electron distribution function $f_e (x,p,t) = f_1$ 
	\begin{equation}
			\label{vlasov_e}
			\frac {\partial f_e}{ \partial t} + \frac{p}{m_e\gamma} \frac{\partial f_e}{\partial x} - eE\frac{\partial f_e}{\partial p} = 0,
		\end{equation}
	the Vlasov equation for the proton distribution function $f_p (x,p,t) = f_2 + f_3 +f_4$
		\begin{equation}
			\label{vlasov_p}
			\frac {\partial f_p}{ \partial t} + \frac{p}{m_p\gamma} \frac{\partial f_p}{\partial x} + eE\frac{\partial f_p}{\partial p} = 0,
		\end{equation}
	and Poisson's equation for the electric field
		\begin{equation}
			\label{poisson}
			\frac{\partial E}{\partial x} = -\frac{e}{\epsilon_0} \left( \int f_e dp - \int f_p dp \right)
		\end{equation}
	The Vlasov-Poisson system is solved using a parallelised version of 
the code detailed in Arber \& Vann (\cite{Arber02}) and Sircombe et al. 
(\cite{Sircombe05}). This uses a split Eulerian scheme in which the 
distribution functions ($f_e, f_p$) are calculated on a fixed Eulerian 
grid. The solver is split into separate spatial and momentum space updates 
(Cheng \& Knorr \cite{Cheng76}). These updates are one dimensional, constant 
velocity advections carried out using the piecewise parabolic method 
(Colella \& Woodward \cite{Colella84}). \\
\subsection{Initial Conditions}
	Throughout our simulations we adopt a realistic mass ratio, 
$m_p / m_e = M_r$, of $M_r = 1836$ and resolve $L$ by 512 cells in the 
$x-$direction, each with length $\Delta x = \pi v_b / 128 \omega_{p,1}$. 
To accurately resolve the filamented phase space distributions of the 
particles that result from the sideband instability (McClements et al. 
\cite{KenPRL}), we use a momentum-space grid with between 4096 and 16384 
grid points ($N_p$). This ensures that the momentum grid spacing for each 
species, $\Delta p_{x,i}$, is small. Specifically,  $\Delta p_{x,i} < 
0.1 m_i v_{th,i}$ where $v_{th,i}^2 = \kappa T_i / m_i$ 
is the thermal speed of species $i$ and where $\kappa$ and $T_i$  are the 
Boltzmann constant and the temperature of species $i$. \\
We set the temperatures of the four species to $T_1 = 5.4 \times 10^5$ K, 
$T_2 = T_3 = 10 T_1$ and $T_4 = 100 T_2$. We thus obtain $v_{th,1} = 
10^{-2}c$. Each species is described by a Maxwellian momentum distribution 
of the form
\begin{equation}
f_i(p) = C_i \exp \left(\frac{-m_i\left(\gamma(p-p_0) - 1\right)}{T_i} \right)
\label{maxwellian}
\end{equation}
where $p_0$ is the initial momentum offset, zero for species 1 and 2, 
$\pm v_b / (1-v_b^2 / c^2)^{0.5}$ for species 3 and 4 respectively. 
For each species the constant $C_i$ is calculated to ensure that 
$\int f_i dp = n_i$ and that $\int \left(f_2 + f_3 + f_4\right) dp = n_1$, 
so there is no net charge in the system.\\
In order to excite the linear instability, we add to the proton distribution 
(species 2, 3 and 4) a density perturbation at the most unstable wavenumber 
of the form $n' = a \cos{k_u x}$, where $a$ is small, typically of the order 
of 1\% of the background density. The minimum beam speed of $v_b = 0.15c$ we 
examine here corresponds to the slow beam in Dieckmann et al. (\cite{MarkPRL}).
The maximum beam speed $v_b = 0.9c$ equals the maximum beam speed in 
Dieckmann et al. (\cite{MarkPoP2}).

\section{Simulation results}
Simulations use a system of normalised units where time is 
normalised to $\omega_{p,1}^{-1}$, space to $c\omega_{p,1}^{-1}$ and momenta 
to $\gamma m_e c$. It follows that the electric field is normalised to 
$\omega_{p,1} c m_1 / e$. Where appropriate in the figures, we re-normalise 
$x$ in units of $2\pi / k_u$, the most unstable wavelength for a given 
initial beam speed. This ensures that the spacial units are identical for 
all beam speeds.


\subsection{Growth and non-linear Saturation of the Buneman instability}

\begin{figure*}
   	\centering
   	\includegraphics[width=\textwidth]{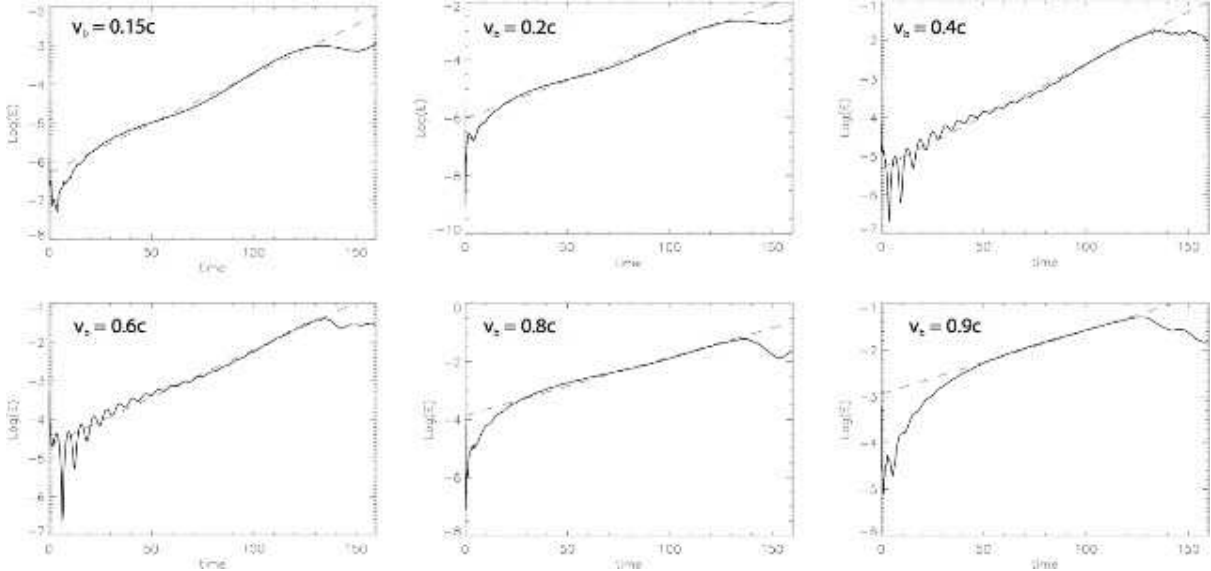}
   	\caption{The logarithmic amplitude of the most unstable mode $E(k_u,t)$ is plotted 
against time, in normalised units, for a variety of beam speeds. The dashed lines on each plot 
represent a linear fit over the region of exponential growth. Using this linear fit we obtain 
growth rates for the relativistic Buneman instability, in normalised units, of $\Omega \approx 
0.0256, 0.0264, 0.0270, 0.0246, 0.0201$ and $0.0137$ for beam speeds of $v_b = 0.15c, 0.2c, 
0.4c, 0.6c, 0.8c$ and $0.9c$, respectively.}
	\label{Fig3}
 \end{figure*}

Figure \ref{Fig3} shows the initial growth stage of the most unstable mode, $E(k=k_u)$, 
for a range of initial beam velocities. From these we estimate the growth rate of the 
instability ($\Omega$), in normalised units, to be  $0.0256, 0.0264, 0.0270, 0.0246, 0.0201$ 
and $0.0137$ for beam speeds of $v_b = 0.15c, 0.2c, 0.4c, 0.6c, 0.8c$ and $0.9c$, respectively. 
These compare favourably with the linear theory. Writing $\alpha(v_b)$ as the ratio between 
observed and theoretical growth rates we find; $\alpha(0.15c) \approx 0.79$, $\alpha(0.2c) 
\approx 0.81$, $\alpha(0.4c) \approx 0.87$, $\alpha(0.6c) \approx 0.87$, $\alpha(0.8c) \approx 
0.86$ and $\alpha(0.9c) \approx 0.72$. As explained in Dieckmann et al. (\cite{MarkPoP2}), 
where a similar systematic reduction has been observed in PIC simulations (in this case by 
15-20\%), this might be connected with the fast growth rate of the instability itself since 
it results in a considerable spread in frequency for the unstable wave. This makes the 
treatment of the instability in terms of single frequencies ($\omega_u$) inaccurate. In truth, 
the growth rate should be lower since the energy of the unstable wave is spread over damped 
frequencies.

\begin{figure}
   	\centering
   	\includegraphics[width=12cm]{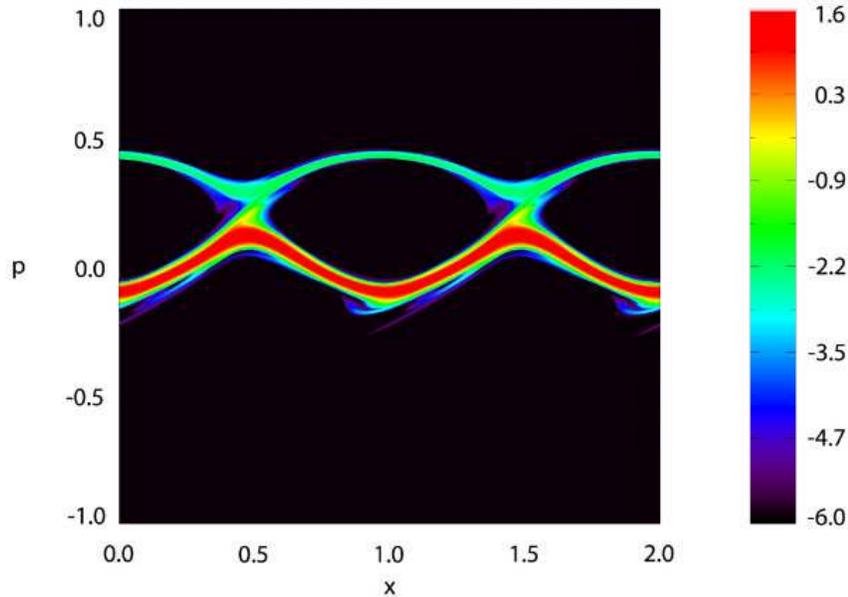}
   	\caption{ Contour plot of $\log (f_e>10^{-6})$ at time $t=150$ for a system 
with an initial beam velocity of $v_b = 0.2c$. The exponential growth of the ESW is in the 
process of saturating via the trapping of electrons, forming a BGK mode at  $p \approx + 0.2.$. 
The thermal spread of beam 2 inhibits its growth slightly, allowing the system to become 
dominated by one ESW. Hence, we do not observe electron trapping around $p \approx - 0.2$. 
Here $x$ is given in units of $2\pi / k_u$.}
	\label{Fig4}
 \end{figure}

\begin{figure}
   	\centering
   	\includegraphics[width=12cm]{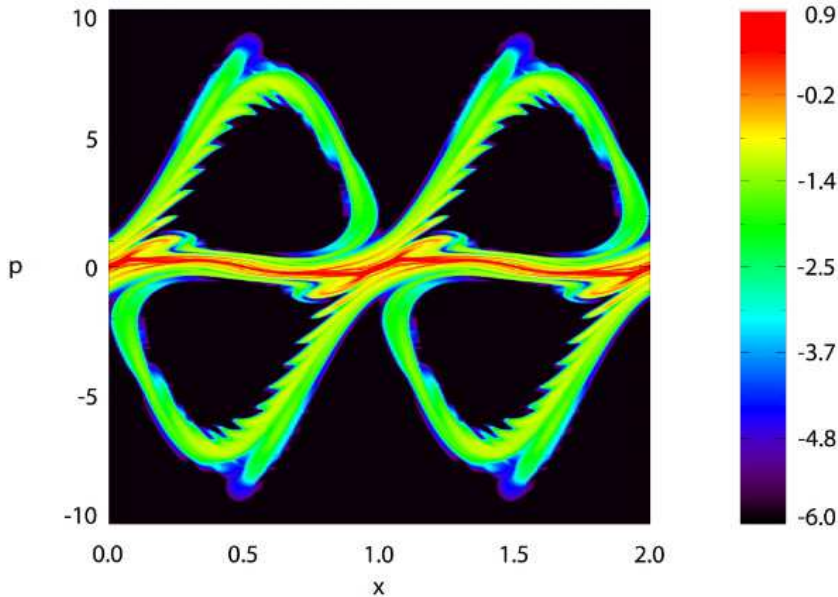}
   	\caption{ Contour plot of $\log (f_e>10^{-6})$ at time $t=200$ for a system with 
a relativistic initial beam velocity of $v_b = 0.9c$. The exponential growth of the ESW is in 
the process of saturating via the trapping of electrons, forming a pair of BGK modes at $p 
\approx \pm 3$. The reduced significance of the higher temperature of species 4, with respect 
to species 3, at high beam speeds allows the growth of forwards and backwards propagating ESWs simultaneously. Here $x$ is 
given in units of $2\pi / k_u$.}
		\label{Fig5}
 \end{figure}

The ESWs saturate by the trapping of electrons and the formation of BGK 
modes. This is shown in Fig. \ref{Fig4}, for the case of $v_b = 0.2c$, 
and Fig. \ref{Fig5} for $v_b = 0.9c$. Here we see the appearance of 
phase-space holes characteristic of particle trapping. While the saturation 
mechanism is the same in both cases, for the high velocity beam we observe 
the development of two counter-propagating BGK modes. This is because at 
lower beam speeds the ESW instability driven by the cooler beam (at $+v_b$) 
dominates whereas for higher, relativistic, beam speeds forwards and 
backwards propagating ESWs grow in unison. At lower beam speeds the increased 
temperature of beam 2 (species 4) is more significant, since the thermal 
velocity of the beam is a larger proportion of the beam speed than is the 
case at $v_b = 0.9c$. Thus, the growth rate for the hot beam is sufficiently 
reduced for the system to be dominated by the growth of the cooler proton 
beam (species 3). We do observe the growth of a counter propagating ESW which 
begins to trap electrons after $t=150$. However, the final momentum 
distribution is clearly dominated by electrons accelerated by the ESW with 
positive phase velocity.
At $v_b = 0.9c$, the thermal spread of both proton beams is negligible in 
comparison to the beam velocities and we observe the growth and saturation 
of ESWs associated with both beams.

\subsection{Wave collapse and secondary electron trapping}

\begin{figure*}
   \centering
   \includegraphics[width=12cm]{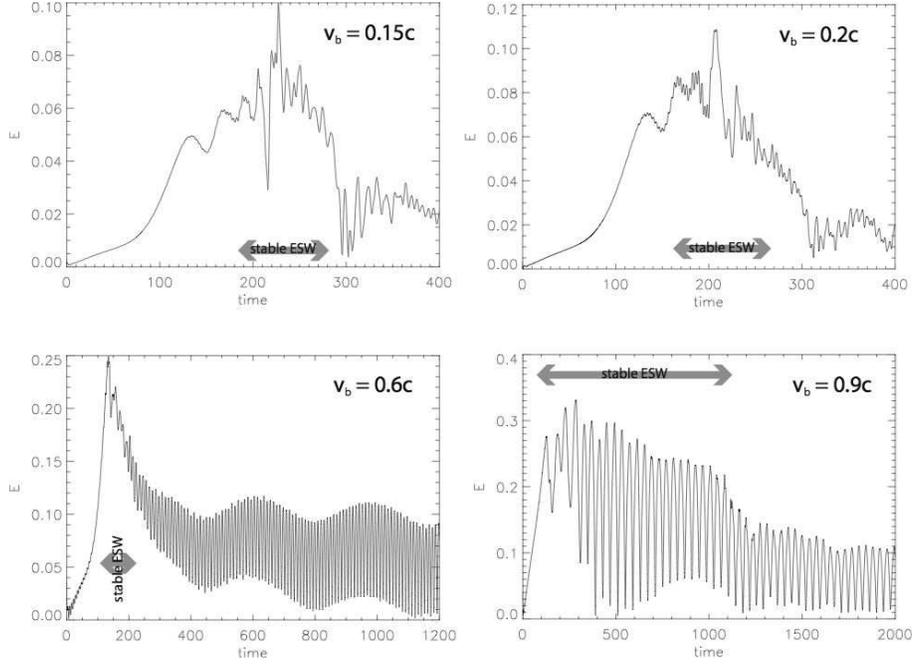}
   \caption{The time evolution of the ESW amplitudes with $k=k_u$ that
   are driven by beam 1 for beam speeds of $0.15c$, $0.2c$,  $0.6c$ and $0.9c$. ESWs generated at 
non-relativistic beam speeds saturate smoothly and stabilise briefly. At $v_b=0.6$ the ESW collapses 
abruptly and at relativistic beam speeds, $v_b=0.9c$ we observe a stabilisation of the ESW for a 
period of almost $1000\omega_{pe}^{-1}$, equivalent to almost $160$ plasma periods. Approximate 
regions of stability are highlighted on each plot.}
   \label{Fig6}
\end{figure*}

Trapping of electrons in electrostatic waves produces BGK modes which eventually collapse via the sideband instability. The sideband instability is due to the nonlinear 
oscillations of the electrons in the potential of the ESW. For electrons
close to the bottom of the wave potential, their oscillation is that of
a harmonic oscillator. The monochromatic bouncing frequency is Doppler
shifted, due to the phase speed of the ESW. These sidebands can couple 
the electron energy to secondary high-frequency ESWs, which must have
a wave number $k \le k_u$ (Krasovsky \cite{Krasovsky}).
The sideband instability is a limiting factor for the lifetime of the
ESW which has implications in the presence of an external
magnetic field in particular. To demonstrate this we show in Fig. \ref{Fig6} the
amplitude of the ESW driven by beam 1 at $k=k_u$ for beam speeds ranging 
from $0.15c$ to $0.9c$.
We find that the ESWs moving at nonrelativistic phase speeds saturate
smoothly, which is in line with the wave saturation of the ESW in the
Vlasov simulation in Dieckmann et al. (\cite{MarkPRL}). Their lifetime
is significant and, in the presence of a weak external magnetic field 
$\vec{B}$ orthogonal to the wave vector $\vec{k}$, the trapped electrons 
would undergo substantial ESA (Eliasson, Dieckmann \& Shukla 
\cite{Eliasson05}). The ESA is proportional to $v_{ph} |\vec{B}|$ and 
the comparatively low $v_{ph}$ would limit the maximum energy the 
electrons can reach.

The ESW driven by the beam with $v_b = 0.6c$ grows to a larger amplitude 
and then collapses abruptly, which confirms the finding in Dieckmann et al. 
(\cite{MarkPRL}) that the BGK modes become more unstable the larger
the ratio between beam speed to electron thermal speed becomes. The Lorentz
force $ev_{ph}|\vec{B}|$ excerted by a $\vec{B} \bot \vec{k}$ would
be significantly higher than for the case $v_b = 0.15$, however the short
lifetime of the wave would here prevent electrons from reaching highly relativistic
speeds. By increasing the beam speed to $v_b = 0.9c$ we obtain a stabilisation
of the ESW, in line with the results in Dieckmann et al. (\cite{MarkPoP2}).
Here the Lorentz force is strong and the lifetime of the saturated ESW is long constituting a formidable electron accelerator, provided
the field evolution is not strongly influenced by $\vec{B}$.

We now summarise the principal results from each beam speed.
Note that in particular the ESWs driven by the mildly relativistic proton 
beams show oscillations after their initial saturation. To identify the 
origin of these fluctuations we apply a Window Fourier Transform to the 
amplitudes of the ESWs with the wavenumbers $k_u / 2 $ and $k_u$. 

\subsection{Beam speed, $v_b = 0.15c$.}

\begin{figure}
   \centering
   \includegraphics[width=12cm]{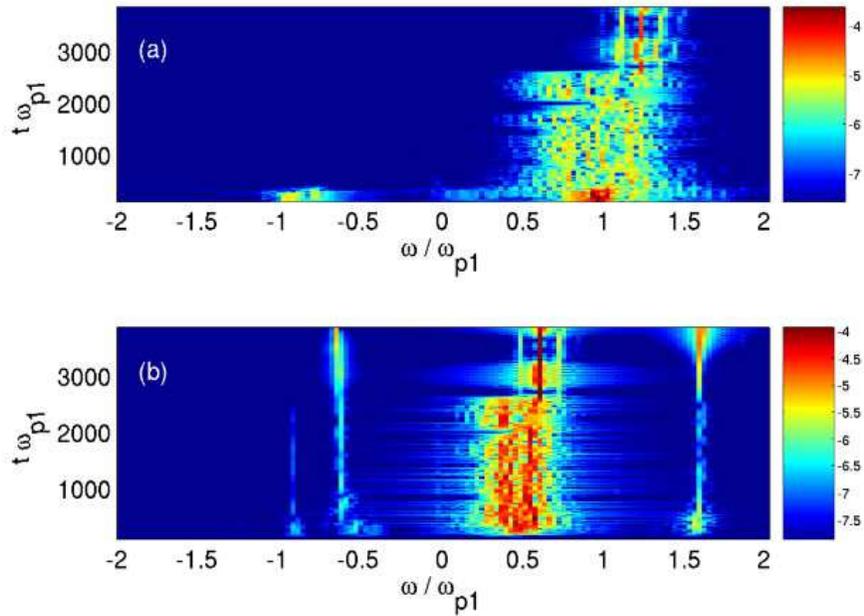}
   \caption{The ESW spectrum for a beam speed of $v_b = 0.15c$:
(a) shows the frequency power spectrum of $k_c=k_u$ and (b) shows the 
frequency power spectrum of $k_c=k_u/2$ as a function of time. The colour 
scale shows $\log_{10} W_2(\omega,t,k_c)$. The ESW driven by beam 1 
with $v_b > 0$ corresponds to the wave with $\omega = \omega_{p,1}$ in (a).}
   \label{Fig7}
   \end{figure}

In Fig. \ref{Fig7} we find a strongly asymmetric ESW growth
for $v_b = 0.15c$. Here the thermal spread of the individual plasma
species is not small compared to $v_b$. Therefore the ESW driven by the cooler
beam 1 grows and saturates first. It rapidly collapses and this
collapse inhibits a further growth of the ESW driven by beam 2.
The initial monochromatic ESW collapses into a broad wave continuum 
centred around $\omega = \omega_{p,1}$. We find the equivalent broad
wave continuum for $k = k_u /2$ centred at $\omega = \omega_{p,1} /2$.
These waves propagate at phase speeds comparable to $v_b$ and represent 
the structures remaining of the initial BGK mode. The ESW spectrum at 
$k = k_u / 2$ further shows two wave bands that are separated by a
frequency of $\omega_{p,1}$ from the main peak. These two wave 
bands could be pumped by a beat between the turbulent structure 
centred at $\omega = \omega_{p,1}/2$, $k = k_u/2$, the turbulent 
structure centred at $\omega = \omega_{p,1}$ and $k= k_u$ and the
Langmuir wave with $k=k_u / 2$. Evidence 
for this is the correlation between both turbulent structures 
and the wave bands at $t\approx 2500$. At this time most wave power at 
$\omega \approx \omega_{p,1} / 2$ is absorbed at $k \approx k_u / 2$. At the 
same time wave power at $\omega \approx \omega_{p,1}$ is absorbed at $k = k_u$ 
while the power in the wave bands with $\omega \approx \omega_{p,1}/2$ and
$\omega \approx 1.5 \omega_{p,1}$ grows at $k=k_u/2$. The faster of these 
two has a phase speed of $\omega / k \approx 3 v_b = 0.45c$. 

\begin{figure}
   \centering
   \includegraphics[width=12cm]{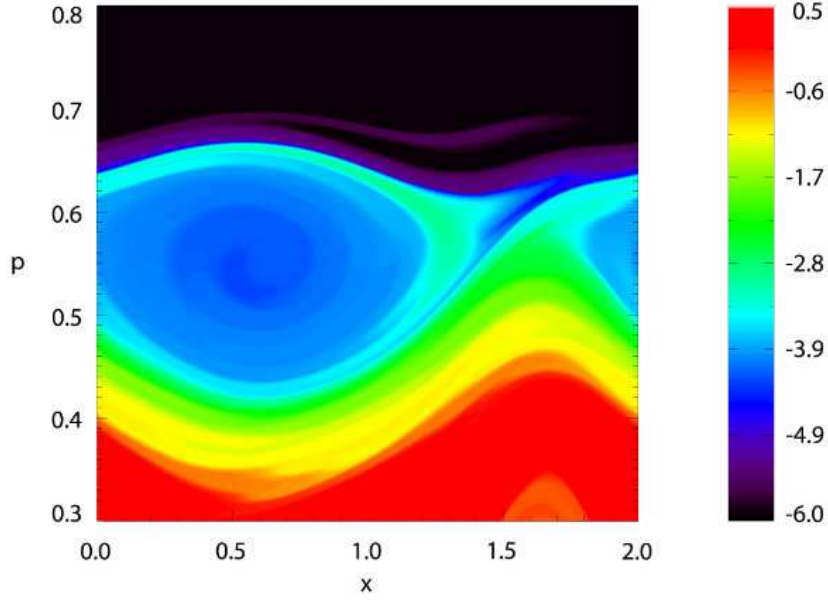}
   \caption{ Contour plot of $\log(f_e>10^{-6})$ for $v_b = 0.15c$ at 
the time $t=3500$: The initial electron BGK modes have collapsed and they can
be shown to form a plateau distribution at $p / m_e c < 0.3$. Centred at 
$p / m_e c \approx 0.52$ we find BGK modes driven by the beat wave.}
   \label{Fig8}
   \end{figure}

This beat wave is thus considerably faster than the initial ESW. By 
its large amplitude it could trap electrons. This is confirmed by 
Fig. \ref{Fig8} where we find a BGK mode in the electron distribution 
centred around a momentum $p / m_e c \approx 0.52$ which corresponds 
to a speed $0.46c$. The fastest electrons of this BGK mode reach 
$p / m_e c \approx 0.68$ or a speed of $0.56c \approx 3.75 v_b$.

\subsection{Beam speed, $v_b = 0.2c$.}
\begin{figure}
   \centering
   \includegraphics[width=12cm]{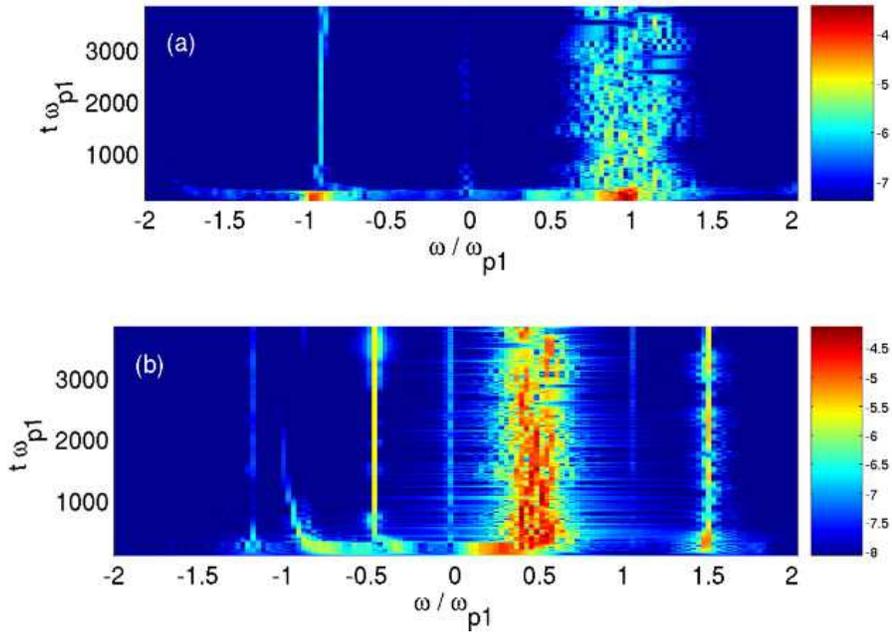}
   \caption{The ESW spectrum for a beam speed of $v_b = 0.2c$:
(a) shows the frequency power spectrum of $k_c=k_u$ and (b) shows
the frequency power spectrum of $k_c=k_u/2$ as a function of time. The 
colour scale shows $\log_{10} W_2(\omega,t,k_c)$. The ESW driven by 
beam 1 with $v_b > 0$ corresponds to the wave with $\omega = \omega_{p,1}$
in (a).}\label{Fig9}
   \end{figure}

The simulation with $v_b = 0.2c$ shows a similar wave coupling as shown in Fig. \ref{Fig9}. We find a turbulent ESW spectrum
that has been driven by the beam 1 and that covers waves with phase
speeds centred at $v_b$ with a spread that is a significant fraction
of the beam speed. Initially an ESW is also driven by beam 2 at 
$\omega \approx \omega_{p,1}$ but, in line with Fig. \ref{Fig7}, it 
collapses simultaneously with the ESW driven by beam 1. The turbulent
wave spectrum with $\omega \approx \omega_{p,1}$ at $k=k_u$ appears
to couple with the equivalent spectrum at $\omega \approx \omega_{p,1}/2$
at $k=k_u / 2$ and the Langmuir wave with $\omega = \pm \omega_{p,1}$ 
and $k=k_u / 2$ to give wave bands at $\omega \approx -0.5 \omega_{p,1}$
and at $\omega \approx 1.5 \omega_{p,1}$. The phase speed of the ESW
band at $\omega \approx 1.5 \omega_{p,1}$ is $\omega / k \approx 3 v_b$
and that of the band at $\omega \approx \omega_{p,1}/2$ is $\omega / k
\approx v_b$. 

\begin{figure}
   \centering
   \includegraphics[width=12cm]{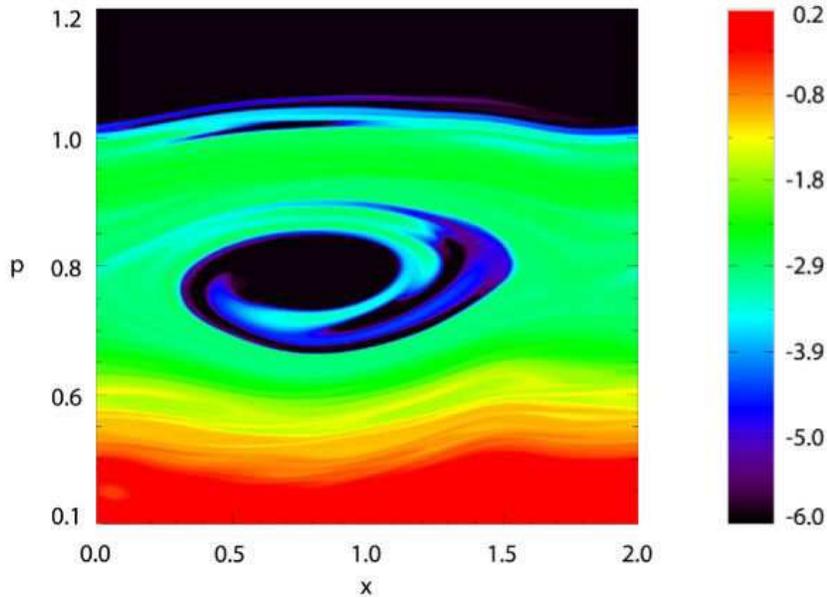}
   \caption{ Contour plot of $\log(f_e>10^{-6})$ for $v_b = 0.2c$ at the 
time $t=2000$: The initial electron BGK modes have collapsed and form a 
plateau distribution. Centred at $p / m_e c \approx 0.75$ we find BGK 
modes driven by the beat wave.}
   \label{Fig10}
   \end{figure}

The large amplitudes of both bands suggests that they might also be 
trapping electrons. This is confirmed by Fig. \ref{Fig10} where we 
show the electron momentum distribution at $t=2000$. We find  BGK modes 
centred at $p / m_e c \approx 0.75$ or a speed of 0.6c. The 
BGK modes extend up to a peak momentum of $p / m_e c \approx 1$ or
a speed of 0.87c respectively equivalent to $4.5 v_b$.

\subsection{Beam speed, $v_b = 0.4c$.}
\begin{figure}
   \centering
   \includegraphics[width=12cm]{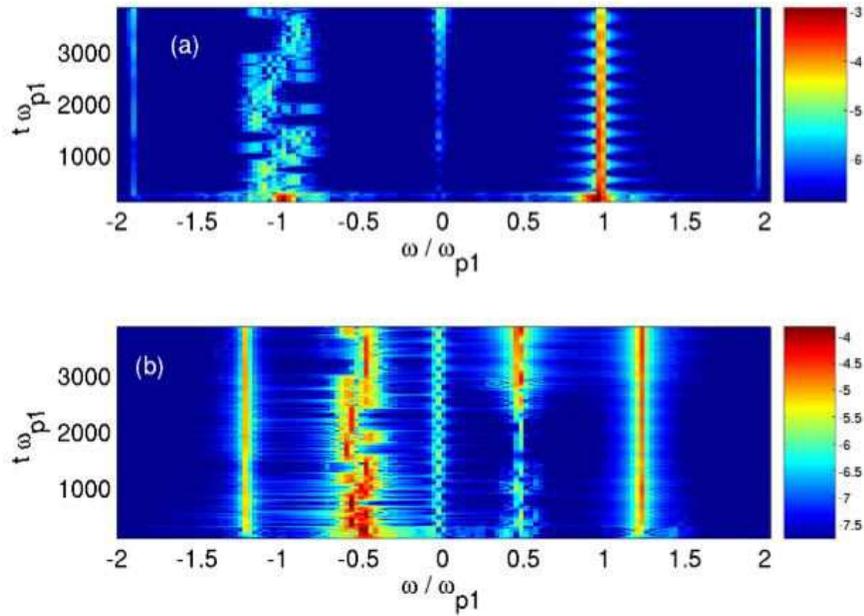}
   \caption{The ESW spectrum for a beam speed of $v_b = 0.4c$:
(a) shows the frequency power spectrum of $k_c=k_u$ and (b) shows the 
frequency power spectrum of $k_c=k_u/2$ as a function of time. The 
colour scale shows $\log_{10} W_2(\omega,t,k_c)$. The ESW driven
by beam 1 with $v_b > 0$ corresponds to the wave with $\omega = 
\omega_{p,1}$ in (a).}\label{Fig11}
   \end{figure}

As we increase the beam speed to $v_b = 0.4c$ the ESW driven by beam 1
stabilises and the collapsing ESW driven by beam 2 drives the
broadband turbulence as we see from Fig. \ref{Fig11}. The peak energy 
of the two counter-propagating ESWs is now comparable. Since we keep
the thermal spread of the beams constant while increasing $v_b$, the
thermal effects are reduced and the growth rate of both waves 
approaches the peak growth rate for the cold beam instability. We find the growth of ESWs at $k=k_u$ 
with $|\omega| \approx 2\omega_{p,1}$. Since these ESWs have the same
wave number as the initial wave they can not be produced by 
self-interaction of the initial ESW since this would also double the
wave number. Instead we believe that these high frequency waves 
correspond to the Doppler shifted bouncing frequency of the trapped
electrons in the potential of the strong ESW. The fastest electrons
reach approximately twice the phase speed $v_{ph}$ of the ESW as
can be seen, for example, at the lower beam speed of $v_b = 0.2c$ in Fig. \ref{Fig4}.
These electrons are thus, due to the fixed wave number $k_u$ of the BGK 
modes, interacting with secondary ESWs with $\omega \approx 2\omega_{p,1}$.
The ESWs with $|\omega| \approx 1.2 \omega_{p,1}$ at $k = k_u/2$ are,
on the other hand, possibly due to a beat between the ESWs with
$|\omega| = \omega_{p,1}$ at $k=k_u$ and the ESWs with $|\omega|=
\omega_{p,1}/2$ at $k=k_u / 2$ and Langmuir waves, as for the slower
ESWs.

\subsection{Beam speed, $v_b = 0.6c$.}
The ESW wave spectrum for $v_b = 0.6c$ is similar to that for $v_b = 0.4c$.
Again we find that both beams grow strong ESWs close to $|\omega_u| = 
\omega_{p,1}$ but that it is the still stronger ESW driven by beam 1 that 
stabilises while that driven by beam 2 collapses into a broadband spectrum. 
This is also reflected by the ESW spectrum at $k_u / 2$ where we find the
strongest wave activity at $\omega \approx -\omega_{p,1}/2$. 
Two additional wave bands with frequencies $\omega \approx \omega_{p,1}$ 
are observed at $k_u / 2$, i.e. long Langmuir waves are produced by the
nonlinear processes. These waves have twice the phase speed of the initial 
ESWs and by their superluminal phase speed they can not interact resonantly 
with the electrons. At $k=k_u$, on the other hand, we find high-frequency 
modes at $|\omega| \approx 1.6 \omega_{p,1}$ with a subluminal
phase speed $\omega / k_u \approx 0.95c$.

   \begin{figure}
   \centering
   \includegraphics[width=12cm]{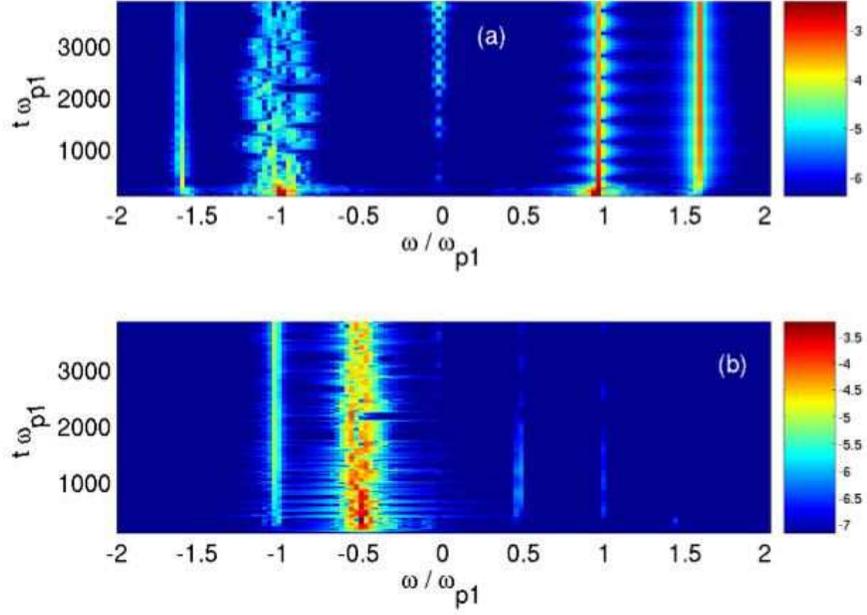}
   \caption{The ESW spectrum for a beam speed of $v_b = 0.6c$:
(a) shows the frequency power spectrum of $k_c=k_u$ and (b) shows the 
frequency power spectrum of $k_c=k_u/2$ as a function of time. The 
colour scale shows $\log_{10} W_2(\omega,t,k_c)$. The ESW driven
by beam 1 with $v_b > 0$ corresponds to the wave with $\omega = 
\omega_{p,1}$ in (a).}\label{Fig12}
   \end{figure}

\begin{figure}
   \centering
   \includegraphics[width=12cm]{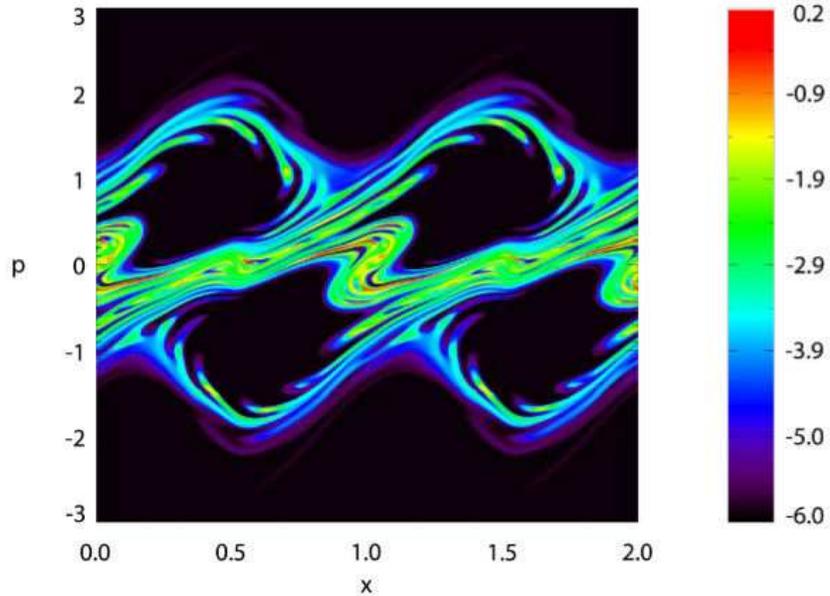}
   \caption{ Contour plot of $\log(f_e>10^{-6})$ for $v_b = 0.6c$ at the 
(early) time $t=200$: The initial electron BGK modes have just been formed 
and the trapped electrons reach a momentum up to $p / m_e c \approx 3$.}
   \label{Fig13}
   \end{figure}

The phase speeds of these ESWs is higher than the peak speed the 
electrons reach in the inital BGK mode as shown in Fig. \ref{Fig13}. 
We can thus not explain its growth by a streaming instability between 
the trapped electrons and, for example, the untrapped electrons. We can
obtain phase speeds of the ESW bands that are higher than the speed of
the trapped electron beam, however, by applying the relativistic Doppler 
shift to the electron bouncing frequency in the ESW wave potential. 
The rest frame of the ESWs moves with the speed $v_{ph} = 0.6c$. The
frequency of the ESWs in the observer frame is, according to Fig. 
\ref{Fig12}, $\omega_o \approx 1.6 \omega_{p,1}$. With the relativistic 
Doppler equation we would obtain a bouncing frequency of the electrons 
in the ESW frame of reference of 
$\omega_b = {([1-v_{ph}/c]/[1+v_{ph}/c])}^{1/2} \omega_o = 0.8 \omega_{p,1}$.
We use the nonrelativistic estimate of the electron bouncing frequency
in a parabolic electrostatic potential $\omega_b^2 = e k_u E / m_e$
and the corresponding width of the trapped electron island 
$v_{tr}^2 = 2eE/m_e k_u$ to eliminate the electric field $E$. We
obtain the relation $\omega_b / k_u = v_{tr}/\sqrt{2}$. Since for
our cold plasma species the velocity width of the island of trapped 
electrons must be comparable to $v_{ph}$ to trap the bulk electrons
we get an estimate for $\omega_b \approx \omega_{p,1} / \sqrt{2}$
which is close to $\omega_o$. We may thus indeed interpret the two 
sidebands observed in Fig. \ref{Fig12} as the Doppler shifted 
bouncing frequency of the electrons. 

\begin{figure}
   \centering
   \includegraphics[width=12cm]{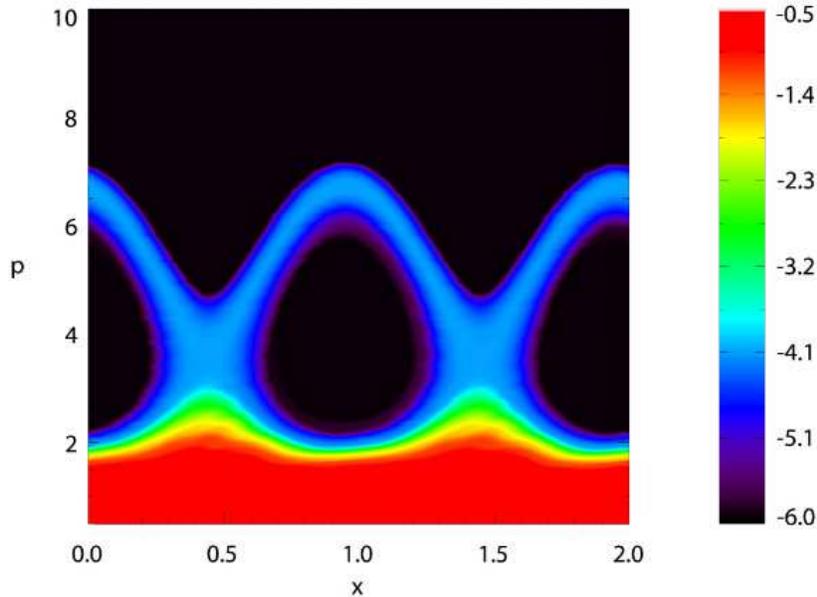}
   \caption{The electron momentum distribution for $v_b = 0.6c$ at the 
(late) time $t=3500$: The initial electron BGK modes have collapsed and 
the electrons that are trapped by the sideband mode reach a momentum 
up to $p / m_e c \approx 7$.}
   \label{Fig14}
   \end{figure}

These sideband modes driven by the beams of trapped electrons have a phase 
speed that is just below $c$ and a large amplitude. Since their phase
speed is comparable to the fastest speed the electrons reach in the initial
BGK mode, they should be capable of trapping some of these electrons.
This is confirmed by Fig. \ref{Fig14} where we find BGK
modes centred at the momentum $p / m_e c \approx 3.4$ which accelerate 
electrons up to the peak momentum $p / m_e c \approx 7$ or a speed of $0.99c$.

 \subsection{Beam speed, $v_b = 0.8c$ and $v_b = 0.9c$.}
As we increase the beam speed to $v_b = 0.8c$ the qualitative evolution
of the ESWs changes. At this high beam speed the thermal spread of the
plasma species is negligible. The growth rates of the waves for both 
beams is approximately that of the cold beam instability and the ESWs grow 
symmetrically. Both saturated ESWs are stable. After the saturation each 
ESW shows a sideband, similar to that in Fig. \ref{Fig13} at a frequency 
modulus $\omega \approx 1.3 \omega_p$. The phase speeds of these sideband
modes are presumably just above $c$. We find a growing second mode at 
$k = k_u /2$ at the frequency modulus $\omega \approx 0.8 \omega_{p,1}$ 
which moves at a superluminal phase speed. 

   \begin{figure}
   \centering
   \includegraphics[width=12cm]{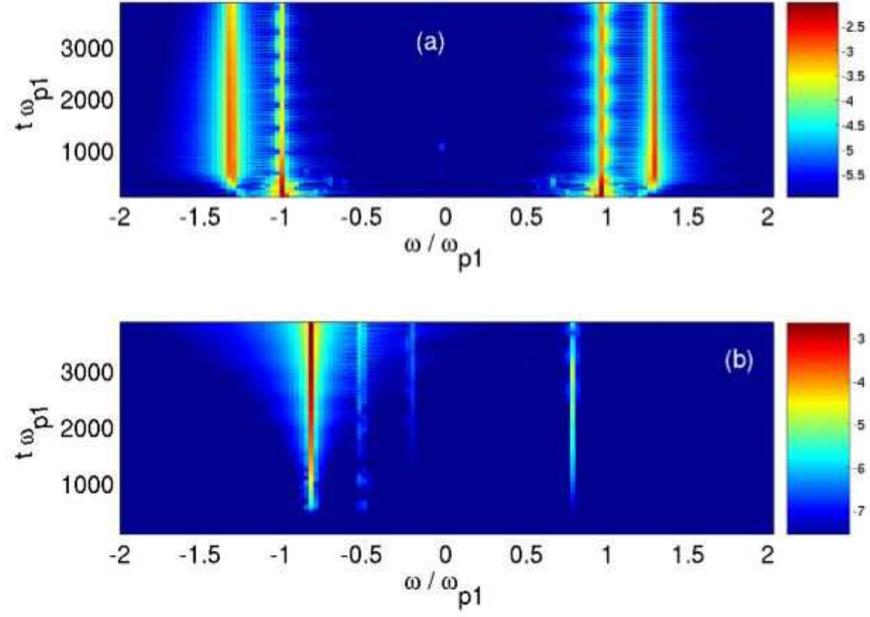}
   \caption{The ESW spectrum for a beam speed of $v_b = 0.8c$:
(a) shows the frequency power spectrum of $k_c=k_u$ and (b) shows the 
frequency power spectrum of $k_c=k_u/2$ as a function of time. The 
colour scale shows $\log_{10} W_2(\omega,t,k_c)$. The ESW driven by 
beam 1 with $v_b > 0$ corresponds to the wave with $\omega=\omega_{p,1}$
in (a).}\label{Fig15}
   \end{figure}

\begin{figure}
   \centering
   \includegraphics[width=12cm]{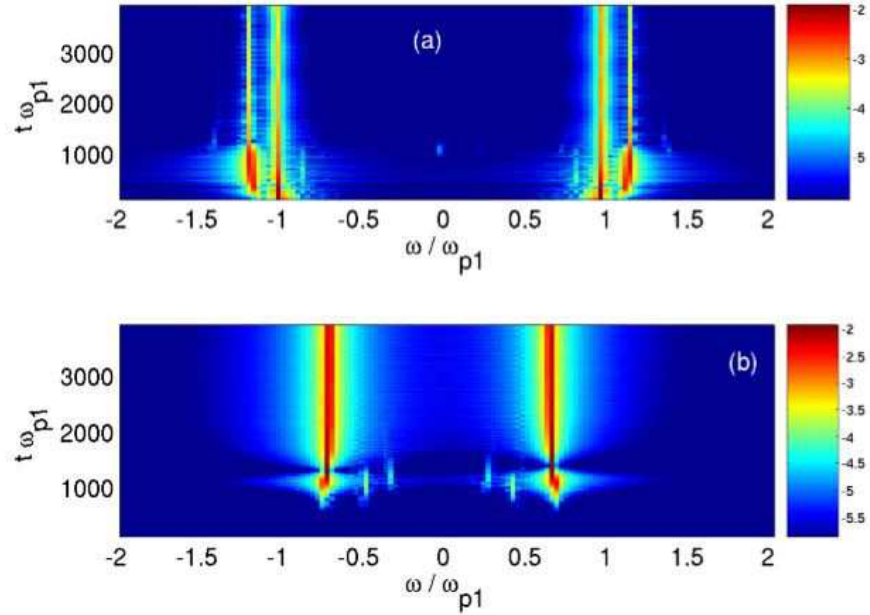}
   \caption{The ESW spectrum for a beam speed of $v_b = 0.9c$:
(a) shows the frequency power spectrum of $k_c=k_u$ and (b) shows the 
frequency power spectrum of $k_c=k_u/2$ as a function of time. The 
colour scale shows $\log_{10} W_2(\omega,t,k_c)$. The ESW driven
by beam 1 with $v_b > 0$ corresponds to the wave with $\omega = \omega_{p,1}$
in (b).}\label{Fig16}
   \end{figure}

We observe the same growth of sideband modes for the fastest beam speed
of $v_b = 0.9c$ in Fig. \ref{Fig16}. The sideband modes at $k=k_u$ have 
a frequency modulus $\omega \approx 1.2 \omega_{p,1}$ and phase 
speeds just above $c$. As for $v_b = 0.8c$ the sideband
mode at $k = k_u / 2$ has a frequency of $\omega \approx 0.8 \omega_{p,1}$ and thus
a superluminal phase speed.

\begin{figure}
   \centering
   \includegraphics[width=12cm]{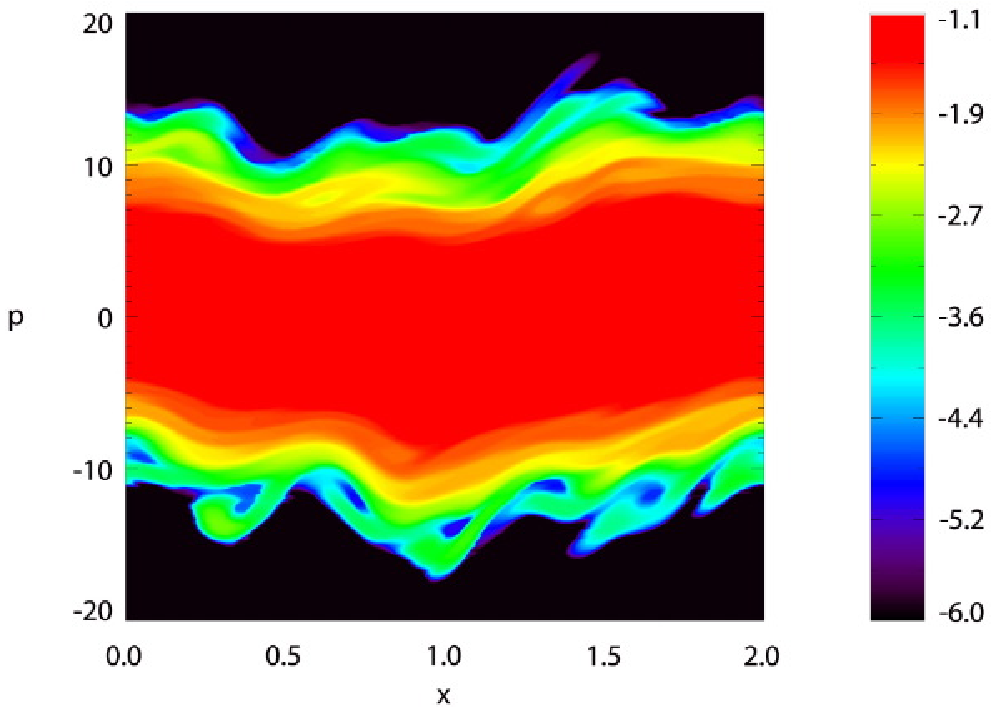}
   \caption{ Contour plot of $\log(f_e>10^{-6})$ for $v_b = 0.9c$ at the 
time $t=4000$: No well-defined BGK mode is visible in the electron distribution
since the sideband unstable modes have a superluminal phase speed.}
   \label{Fig17}
   \end{figure}

For both beam speeds $v_b = 0.8c$ and $v_b = 0.9c$ the sideband modes
appear to have a superluminal phase speed and they can therefore not
trap electrons. No secondary BGK modes should develop for these 
beam speeds.
This is confirmed by Fig. \ref{Fig17} where we show the phase space
distribution of the electrons for $v_b = 0.9c$ at the simulation's end.
The phase space distribution at high momenta shows no evidence of a
BGK mode despite the strong sideband modes in Fig. \ref{Fig16}.

In contrast to the PIC simulations in Dieckmann et al. (\cite{MarkPoP2})
the Vlasov simulation code shows the growth of sideband modes and what
appears to be waves resulting from a parametric instability. These
secondary waves grow to a large amplitude at which they can nonlinearly
interact with the electrons. For the beam speeds up to $v_b = 0.4c$
the waves generated by the parametric interaction have been strongest.
Here the turbulent wave fields interact with the waves with $\omega_{p,1}$
to produce a wave with a higher frequency. In contrast to plasma beat
wave accelerators, which have recently been reviewed by Bingham, Mendonca 
\& Shukla (\cite{Bingham04}), for which two high-frequency electromagnetic 
waves beat 
to yield a low frequency ESW that can accelerate the electrons, our parametric
coupling couples low frequency ESWs to an ESW with a higher frequency.
Its larger phase speed can accelerate the trapped electrons to higher peak 
speeds. We may thus call it the ``inverse plasma beat wave accelerator''.
For a beam speed of $v_b = 0.6c$ the turbulent wave fields do not 
noticably interact with $\omega_{p,1}$. Instead a sideband unstable mode
with $k=k_u$ develops, i.e. at the largest allowed wave number for
nonrelativistic BGK modes (Krasovsky \cite{Krasovsky}). This mode has
for $v_b \le 0.6c$ a phase speed below $c$ and the electron phase space 
distribution shows a fast BGK mode driven by it. For even higher
$v_b$ the probably superluminal phase speed of the sideband modes
suppresses their interaction with the electrons.

\begin{figure}
   \centering
   \includegraphics[width=12cm]{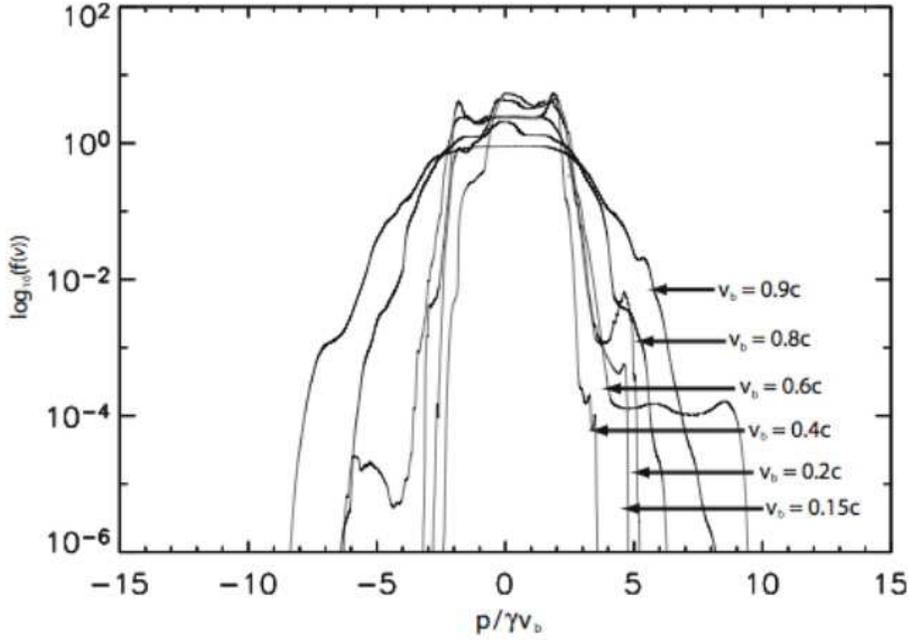}
   \caption{The electron momentum distributions at the simulation's end
times: }
   \label{Fig18}
   \end{figure}

The complex spectrum of secondary waves and their nonlinear interactions
with the electrons, which has not been observed clearly by Dieckmann
et al. (\cite{MarkPoP2}), suggests a stronger dependence of the electron
heating on $v_b$ than for the simulations by Dieckmann et al. 
(\cite{MarkPoP2}). There the momentum distributions could be matched
if one were to scale the momentum axis to $v_b \gamma (v_b)$. We thus
do the same here and we integrate the electron phase space distribution
over $x$. The result is shown in Fig. \ref{Fig18}.

We find that the final momentum distributions in the chosen normalization
of the p-axis agree well up to $v_b = 0.6c$ and for densities larger
than $10^{-2}$. Here the Vlasov code results are similar to the equivalent
PIC simulations in Dieckmann et al. (\cite{MarkPoP2}). We find, however,
significant differences at lower densities, which are not well-represented
by the PIC code, and at the corresponding higher momenta. The parametric 
instabilites and the sideband instabilities
have further accelerated electrons beyond the peak momenta measured in
Dieckmann et al. (\cite{MarkPoP2}). This is particularly pronounced
for the beam speed $v_b = 0.6$ in which we find an electron density
plateau extending to a value of +10 for the normalised momentum, i. e.
twice as high as the corresponding value at negative momenta. For these
beam speeds the secondary waves have been most efficient as electron
accelerators. A further increase of $v_b$ beyond 0.6c yields broadening
momentum distributions. The peak momentum in units of $m_e c$ is, for
positive momenta comparable for $v_b  = 0.6c$ and for $v_b = 0.8c$.
The peak momentum for $v_b = 0.9c$ is about twice as high as for
$v_b = 0.6c$. 

The peak relativistic kinetic energies $K = mc^2 e^{-1}(\gamma - 1)$ 
in eV the electrons reach are $K (v_b = 0.15c) = 1.3 \times 10^5$ eV,
$K(v_b = 0.2c) = 2.2 \times 10^5$ eV, $K(v_b = 0.4c) = 5\times 10^5$ eV,
$K(v_b = 0.6c) = 3.2 \times 10^6$ eV, $K(v_b = 0.8c) = 3.6 \times 10^6$ eV
and $K (v_b = 0.9c) = 8 \times 10^6$ eV. Note that all these peak
electron energies are comparable or above the threshold energy of
$10^5$ eV: the injection energy for Fermi acceleration at perpendicular shocks, given by Treumann \& Terasawa (\cite{Treumann}). 

\section{Discussion}
	
The observation of the emission of highly energetic cosmic ray particles by SNRs suggests the acceleration of particles from the thermal pool of the ISM plasma to highly
relativistic energies by such objects. The acceleration site is apparently linked to the shock that
develops as the supernova blast shell encounters the ambient plasma (Lazendic \cite{Lazendic04}). Such shocks
are believed to accelerate electrons and ions to highly relativistic energies by
means of Fermi acceleration (Fermi \cite{Fermi49}; Fermi \cite{Fermi54}). The
Fermi acceleration of electrons is most efficient if the shock is quasi-perpendicular
(Galeev \cite{Galeev84}). For such shocks, however, Fermi acceleration works only 
if we find a relativistically hot electron population prior to the shock encounter, 
since slow electrons could not repeatedly cross the shock and pick up energy. Since 
the plasma, into which the SNR shock expands, has a thermal speed comparable to that 
of the ISM or the stellar wind of the progenitor star with temperatures of up to a few eV, if we take the solar wind as reference, no mildly relativistic 
electrons may exist. A mechanism is thus required that accelerates electrons up to 
speeds at which their Larmor radius exceeds the shock thickness.
As Galeev (\cite{Galeev84}) proposed, electrons could be pre-accelerated from the 
initial thermal pool to mildly relativistic energies by their interaction with 
strong ESWs in the foreshock region which, in turn, are driven by shock-reflected 
beams of ions. 

We have examined in this work the growth, saturation and collapse 
of ESWs in a system dominated by the presence of two counter-propagating proton 
beams. The currents of both beams cancel, allowing the introduction of periodic 
boundary conditions. These initial conditions can be motivated 
as follows: The upstream protons, that have initially been reflected by the shock, are rotated by the global magnetic 
field oriented perpendicularly to the shock normal and return as a second 
counter propagating proton beam, with a slightly increased temperature. 
The system modelled in this work represents a small region ahead of a SNR shock. {Here the magnetic field has been neglected in order to focus on the ESWs 
and nonlinear BGK modes. 
While this does not describe a complete model for the foreshock dynamics of high 
Mach number shocks, it is applicable to parts of the foreshock of perpendicular shocks where local variations magnetic field cause it to vanish, or become beam-aligned. These variations may, for example, be} due to the 
presence of turbulent magnetic field structures in the foreshock of SNR shocks 
(Jun \& Jones \cite{Jun99}; Lazendic et al. \cite{Lazendic04}).

The system is unstable to the relativistic Buneman instability (Thode \& Sudan \cite{Thode}) 
which saturates via the trapping of electrons to form BGK modes (Rosenzweig \cite{Rosenzweig}). 
These trapped particle distributions are themselves unstable to the sideband instability and 
collapse after a period of stability. 

{Since our model assumes that the particle trajectories are not 
affected by the local magnetic field and because we do not consider 
here electromagnetic waves or waves in magnetised plasma, we are able to utilise} a relativistic, 
electrostatic Vlasov code. Previous work, for example (Thode \& Sudan \cite{Thode}; Dieckmann et al. 
\cite{MarkPoP1}; Dieckmann et al. \cite{MarkPRL}; Shimada \& Hoshino \cite{Shimada1}) has made 
extensive use of PIC codes for this problem and in particular Dieckmann et al. (\cite{MarkPRL}) 
have compared Vlasov and PIC codes in certain conditions. We benefit from the Eulerian Vlasov code's 
ability to resolve electron and ion phase space accurately, irrespective of the local particle density. 
This allows us to identify a secondary acceleration mechanism which may not be immediately apparent 
otherwise.
Overall the results of this work are in agreement with previous 
studies (Dieckmann et al. \cite{MarkPoP1}; Dieckmann et al. \cite{MarkPRL}) showing the lifetimes 
of the BGK modes to be dependent on the initial beam velocity.  As $v_b$ is increased, we observe a 
reduced ESW stability up to $v_b = 0.8c$ but with a significantly increased stability at the highest 
beam speed. 
	
At low beam velocities ($v_b = 0.15c, 0.2c, 0.4c, 0.6c$) we observe long wavelength, high phase-velocity 
modes. These are able to trap electrons, producing a population with kinetic energies above the injection 
energy for Fermi acceleration, even at non-relativistic beam, and thus shock, velocities. We believe these secondary 
(that is to say, not associated with the initial ESW saturation) trapped distributions to be the result 
of trapping in ESWs produced by parametric coupling between low frequency oscillations and plasma waves 
at $\omega_{p,1}$. Above $v_b=0.6c$ the ESWs produced by this coupling have super-luminal phase velocities 
and are unable to trap electrons. Hence we do not observe such BGK modes in the case of $v_b = 0.8c,$ or 
$0.9c$. Our simulation box, at $4\pi / k_u$ in length, can only accommodate one wavemode with wavenumber 
below that of the most unstable mode and this may have an influence on the appearance of this secondary 
acceleration. 

Future work has to examine how these parametric instabilities depend on a 
magnetic field and on the introduction of a second spatial dimension.
We will need to consider significantly larger simulation boxes, capable of 
resolving a broader spectrum below $k_u$. It may be the case that the availability of modes with $k<k_u$ 
will result in the partition of ESW energy across a greater region of the spectrum, perhaps inhibiting 
the trapping of electrons at high velocity. However, it may be the case that our observed coupling and 
resultant electron trapping is the first step of a cascade, capable of accelerating electrons to high 
energies for relatively modest shock velocities. This is since, even for $v_b$ as low as 0.15c or 
corresponding shock speeds of $7.5 \times 10^{-2}c$ which can be reached by the fastest SNR main shocks 
(Kulkarni et al. \cite{Kulkarni1998}), electrons can reach energies of $10^5$ eV. According to 
Treumann \& Terasawa (\cite{Treumann}), this may increase the electron gyroradius beyond the shock thickness 
by which they can repeatedly cross the shock front. These repeated shock crossings allow the electrons 
to undergo Fermi acceleration to highly relativistic speeds. Even higher energies could be achieved if 
we were to get shock precursors that outran the main shock as has been observed for the supernova SN1998bw 
(Kulkarni et al. \cite{Kulkarni1998}). The numerical simulations in this work thus present strong 
evidence for the ability of ESWs and processes driven by electrostatic turbulence to accelerate electrons beyond the threshold 
energy at which they can undergo Fermi acceleration as it has previously been 
proposed, for example 
by Galeev (\cite{Galeev84}). 

\section{Acknowledgements}

This work was supported in part by: the European Commission through
the Grant No. HPRN-CT-2001-00314; the Engineering and Physical
Sciences Research Council (EPSRC); the German Research Foundation (DFG); and the United Kingdom Atomic
Energy Authority (UKAEA). The authors thank the Centre for Scientific Computing (CSC) at the University of Warwick, with support from Science Research Investment Fund grant (grant code TBA), for the provision of  computer time.
N J Sircombe would like to thank Padma Shukla and the rest of the Institut f\"ur Theoretische Physik IV at the Ruhr-Universit\"at Bochum for their kind hospitality during his stay.

 \end{document}